\documentclass[%
reprint,
superscriptaddress,
showpacs,
amsmath,amssymb,
aps,
prl,
longbibliography,
]{revtex4-2}

\usepackage{psfrag,graphicx,epsfig,color}
\usepackage{dcolumn}
\usepackage{bm}
\usepackage{natbib}
\usepackage{float}
\usepackage{xcolor}
\usepackage{subfigure}
\usepackage{nicefrac}
\usepackage{multirow}

\def\re    {Re_\lambda}
\def\uu {{\mathbf{u}}}

\def\ww {{\boldsymbol{\omega}}}

\definecolor{mygreen}{rgb}{0,0.7,0.}

\newcolumntype{C}[1]{>{\centering\arraybackslash}p{#1}}

\begin{document}

\title{
Unified multifractal description of 
longitudinal and transverse intermittency  \\
in fully developed turbulence
}
\author{Dhawal Buaria }
\email[]{dhawal.buaria@ttu.edu}
\affiliation{Department of Mechanical Engineering, Texas Tech University, Lubbock, TX 79409}
%


%

%

\date{\today}

\begin{abstract}

Small-scale intermittency is a defining feature of fully developed fluid turbulence, 
marked by rare and extreme fluctuations of velocity increments and gradients 
that defy mean-field
descriptions. Existing multifractal descriptions of intermittency
focus primarily on longitudinal increments and gradients,
despite mounting evidence that transverse components exhibit distinct and 
stronger intermittency. 
Here, we develop a unified multifractal framework that 
jointly prescribes longitudinal and transverse velocity increments,
and extends to gradients. We derive explicit relations
linking inertial-range scaling exponents of structure functions to moments
of velocity gradients in dissipation range. Our results reveal that longitudinal
gradient scaling is solely prescribed by longitudinal structure functions, 
as traditionally expected; however, transverse gradient scaling is
prescribed by mixed longitudinal-transverse structure functions.
Validation with high-resolution direct numerical simulations of isotropic turbulence,
at Taylor-scale Reynolds number up to $1300$ demonstrates
excellent agreement,
paving way for a more complete and predictive description of intermittency faithful
to the underlying turbulence dynamics.

\end{abstract}

\maketitle



Intermittency, the spontaneous emergence of rare, 
intense fluctuations interspersed within otherwise 
quiescent dynamics, is a hallmark of complex
far-from-equilibrium systems 
\cite{Sreeni97, bruno2013solar, matthaeus2015, majumdar2020extreme}. 
From a statistical standpoint, it is characterized by pronounced 
departures from Gaussianity 
emanating from a breakdown of simple self-similarity arguments, 
leading to anomalous scaling laws and dominance
of extreme events in high-order moments \cite{halperin1969, paladin87an}. 
Fluid turbulence provides a paradigmatic example,
where intermittency naturally manifests 
as energy is transferred from large to small scales,
engendering rare and intense fluctuations
of velocity increments and gradients at the small scales
\cite{Frisch95, Sreeni97}.
These events, playing a crucial role in 
in dissipation, mixing
and transport, 
invalidate the mean-field description 
proposed in the seminal work of 
Kolmogorov (1941) \cite{K41a}, hereafter K41.
Despite decades of progress, their quantitative 
characterization remains incomplete and a 
central obstacle to a universal description of small-scale turbulence
\cite{sreeni25}.

The intermittency of velocity increments may be analyzed
through its longitudinal component:
$\delta u_r = u(x+r) - u(x)$, 
where the velocity component
$u(x)$ is in the direction of the separation $r$;
or the transverse component; 
$\delta v_r = v(x+r) - v(x)$, 
corresponding to velocity component
$v(x)$ orthogonal to separation direction. 
Building upon K41, one expects the 
moments of these increments, 
called structure functions, 
to exhibit anomalous power-law scalings in the inertial-range
$\eta \ll r \ll L$, where $L$ is the large scale 
and $\eta$ is the viscous cutoff scale \cite{Frisch95, Sreeni97}. 
At the smallest scales, these increments naturally 
lead to the longitudinal and transverse velocity gradients,
$\partial_x u = \lim_{r\to0} \delta u_r/r$ and
$\partial_x v = \lim_{r\to0} \delta v_r/r$, respectively,
which directly relate to energy dissipation and enstrophy, 
and characterize intermittency at the viscous scales.

One can loosely equate the longitudinal and transverse quantities
to stretching and rotational motions of turbulent eddies, respectively,
which are intimately coupled through the Navier-Stokes
dynamics and are the key mechanisms driving the energy cascade
\cite{Tsi2009, carbone20, johnson2020}. 
Despite this duality, traditional descriptions of intermittency--most notably 
multifractal models \cite{MS91, SL94, Frisch95}--have focused primarily on
describing longitudinal velocity increments and gradients
\footnote{A reason for this is that only the 1D longitudinal component
along the streamwise direction was accessible in early wind tunnel
experiments. Also, it is arguably easier to pose turbulence
theory in terms of longitudinal increments, since the
the $4/5$-th law, corresponding to its third moment, 
is exactly derivable from Navier-Stokes equations.
In contrast, the $4/3$-th law is also exactly derivable, but requires
both longitudinal and transverse increments \cite{Frisch95}},  
with transverse statistics simply presumed to follow 
identical scaling relations as their longitudinal counterparts. 
However, this viewpoint has become untenable in light of
substantial experimental and numerical evidence 
demonstrating disparate anomalous scaling behaviors
between longitudinal and transverse statistics, 
with the latter being more intermittent
\cite{dhruva97, chen97, shen02, Dubrulle:2019, BPBY2019, iyer2020,
BS_2023, Khurshid_2023}.
These observations underscore the need for a unified
framework capable of capturing both
longitudinal and transverse statistics
for an accurate and complete description of intermittency. 

In this Letter, motivated by the above consideration, 
we develop a joint multifractal framework
which simultaneously prescribes longitudinal
and transverse velocity increments, and extend it to 
velocity gradients. 
We derive generalized relations 
directly relating the scaling exponents of
structure functions in the inertial-range
to those of gradient moments in the dissipation range.
We find that the scaling of longitudinal
gradients is solely prescribed by inertial-range exponents
of longitudinal structure functions, consistent with traditional
single-variable multifractal description. In contrast,
the scaling of transverse gradients is governed not by
transverse structure functions alone, but rather  
mixed longitudinal-transverse structure functions. These 
predictions are validated against 
high-resolution direct
numerical simulations (DNS) of isotropic turbulence,
showing remarkable agreement, and 
establishing a unified, predictive framework
for intermittency that captures
both longitudinal and transverse statistics.

\paragraph{\bf Joint multifractal framework:}

In traditional univariate multifractal description,
the longitudinal velocity increment across a scale $r$ is
taken to be locally H\"older continuous:
$\delta u_r \sim U (r/L)^h$, where the exponent
$h$ varies in the interval $[h_{\rm min}, h_{\rm max}]$.
Here, $U$ represents the large-scale velocity, typically given by 
the rms of velocity fluctuations.
Each value of $h$ is realized on a set with fractal dimension $D(h)$,
also termed the multifractal spectrum, such that the
probability to observe the exponent $h$ at scale $r$
is given as: $P_r(h) \sim (r/L)^{3-D(h)}$.
Thereafter, one can appropriately obtain the scaling of structure
functions and also gradient moments (see e.g. \cite{Frisch95}
for a detailed exposition). 
In principle, a similar
description can be formulated for transverse increment $\delta v_r$
\footnote{One can consider a different and independent multifractal spectrum 
compared to the longitudinal counterpart, or can also assume they are
the same},
but with the drawback that it completely ignores any coupling
between longitudinal and transverse components, which
is known to  play a central role in intermittency 
\cite{BPBY2019, carbone20, johnson2020, BBP2020, BP2022, BPB2022}.  

The above shortcoming can be resolved by considering 
a bivariate multifractal
description \cite{meneveau1990joint}. The essential idea is that 
we jointly consider longitudinal
and transverse increments to be H\"older continuous
\begin{align}
\delta u_r  \sim U (r/L )^{h_1}  \ , 
\quad
\delta v_r  \sim U (r/L )^{h_2} \ ,
\label{eq:urh}
\end{align}
with  exponents $h_1$ and $h_2$
prescribed by a joint multifractal
spectrum $D(h_1, h_2)$ and the corresponding
probability $P_r(h_1, h_2) \sim (r/L)^{3 - D(h_1, h_2)}$. 
Thereafter, the mixed-structure functions of
order $p_1, p_2$ in longitudinal and transverse directions can be obtained
by integrating over the full bivariate distribution
\begin{align}
\begin{split}
&{\langle (\delta u_r)^{p_1} 
(\delta v_r)^{p_2} \rangle}/{ U^{p_1 + p_2}}  \\ 
& \qquad\sim  \int \int  \left( \frac{r}{L} \right)^{p_1 h_1 + p_2 h_2 + 3 - D(h_1, h_2)} dh_1 dh_2  \ . 
\end{split}
\end{align}
From the steepest descent estimation in the limit $r/L \to 0$,
it follows that
\begin{align}
\frac{\langle (\delta u_r)^{p_1} 
(\delta v_r)^{p_2}  \rangle }{ U^{p_1 + p_2}} \sim 
\left( \frac{r}{L} \right)^{\zeta_{p_1,p_2}}  \ , 
\end{align}
with the scaling exponents
given as
\begin{align}
\zeta_{p_1, p_2} &= \inf_{h_1, h_2} \,
[ p_1 h_1 + p_2 h_2 + 3 - D(h_1, h_2) ]  \ .
\label{eq:zetap}
\end{align}
Thus, for a given value of $p_1, p_2$, 
the critical H\"older exponents $h_1^*, h_2^*$ which dictate the scaling 
exponents are obtained by solving the system: 
\begin{align}
\frac{\partial D}{\partial h_1} (h_1^*, h_2^*) = p_1  \ , 
\quad
\frac{\partial D}{\partial h_2} (h_1^*, h_2^*) = p_2  \ .
\label{eq:hcrit}
\end{align}
We note that the work of  \cite{meneveau1990joint} also considered
similar joint multifractal measures, but for
locally-averaged dissipation and enstrophy.
Furthermore, it did not extend to gradients, which is performed next.

\begin{figure*}
\centering
\includegraphics[width=0.95\textwidth]{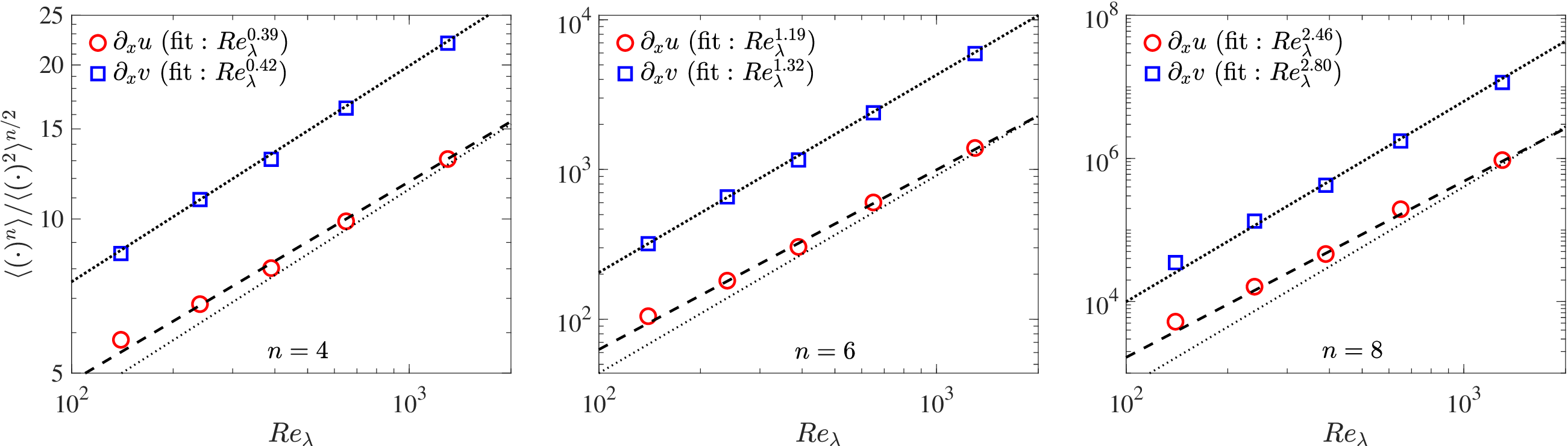} 
\caption{
Central moments of longitudinal and transverse velocity gradients
for moment orders (a) $4$, (b) $6$, and (c) $8$, together with the 
best-fit power-law scalings. 
}
\label{fig:mom_all}
\end{figure*}

\paragraph{\bf Extension to gradients:} 
Combining the definitions of 
longitudinal and transverse 
gradients with Eq.~\eqref{eq:urh} leads to
\begin{align}
\partial_x u \sim  \frac{U}{L} \left( \frac{r}{L} \right)^{h_1 -1} \ \Bigg|_{r=\eta} \ ,
\ \ 
\partial_x v \sim  \frac{U}{L} \left( \frac{r}{L} \right)^{h_2 -1} \ \Bigg|_{r=\eta}  \ , 
\label{eq:jgrad}
\end{align}
where $\eta$ denotes the viscous cutoff scale.
This scale is determined by imposing
that the local Reynolds number of an eddy is
unity \cite{Paladin87, SreeniYakhot:2021}. 
Conventionally, this condition is imposed using
the longitudinal velocity increment: 
\begin{align}
\frac{\delta u_r \  r}{\nu} = 1   \ .
\label{eq:re1}
\end{align}
which combined with Eq.~\eqref{eq:urh}
and evaluated at $r=\eta$, gives 
\begin{align}
\eta/L \sim Re^{-1/(1 + h_1)} \ .
\label{eq:etah}
\end{align}
where we have used $Re=UL/\nu$. 
It is easy to see that the Kolmogorov
length scale $\eta_K = \eta(h_1=\frac{1}{3})$.
It is worth noting that the transverse increment,
and in particular the exponent $h_2$, does not play a role in 
determining $\eta/L$.
This asymmetry is not coincidental, and we will return to it soon.

From Eq.~\eqref{eq:jgrad},
the joint moment of longitudinal and transverse gradients, 
of orders $n_1$ and $n_2$, respectively, can be written as  
\begin{align}
&\frac{\langle (\partial_x u)^{n_1} 
(\partial_x v )^{n_2} \rangle} 
{(U/L)^{n_1 + n_2}}  \\ \nonumber
&\sim  
\int_h \left( \frac{r}{L} \right)^{n_1(h_1-1) + n_2(h_2-1) + 3 - D(h_1, h_2)}  
\ dh_1 \, dh_2  \ \Bigg|_{r=\eta}  \ .    
\end{align}
Substituting the dissipative cutoff from Eq.~\ref{eq:etah}
and invoking the steepest-descent
argument in the limit $Re \to \infty$ 
yields the power-law dependence
\begin{align}
\frac{\langle (\partial_x u)^{n_1} (\partial_x v)^{n_2} \rangle}
{(U/L)^{n_1 + n_2}} \sim 
Re^{\rho_{n_1,n_2}}
\label{eq:dudxn}
\end{align}
where the scaling exponent is given as  
\begin{align}
\rho_{n_1,n_2} = \sup_{h_1,h_2}  \ \frac{n_1(1 - h_1) + n_2(1 - h_2) - 3 + D(h_1, h_2)}{1+h_1}  \ .
\label{eq:rhon}
\end{align}

The results in Eq.~\eqref{eq:zetap}
and Eq.~\eqref{eq:rhon} constitute a natural
generalization of the univariate multifractal model \cite{Frisch95}
to a joint description of longitudinal and transverse fluctuations. 
The knowledge of $D(h_1, h_2)$ uniquely
determines the scaling exponents (of both structure functions and
and gradient moments). Conversely, the
knowledge of any one set of exponents uniquely determines
$D(h_1, h_2)$ through an appropriate Legendre transform, 
and thus fixes the other set as well. 
For instance, the relation in 
Eq.~\eqref{eq:zetap} can be inverted 
to write:
$D(h_1, h_2) = \inf_{p_1, p_2} \, [p_1 h_1 + p_2 h_2 + 3 - \zeta_{p_1, p_2}]$.

To that end, the mixed-gradient exponents $\rho_{n_1, n_2}$
can be directly related to mixed-increment exponents $\zeta_{p_1, p_2}$ without 
explicitly using $D(h_1, h_2)$. 
As shown in the Appendix B, this yields the result 
\begin{align}
\rho_{n_1, n_2} = p_1(n_1, n_2) - n_1 \, .
\label{eq:rho_def}
\end{align}
where the function  $p_1(n_1, n_2)$ is obtained as
the solution of 
\begin{align}
\label{eq:p_def_1}
\zeta_{p_1, p_2} + (p_1 + p_2) &= 2 (n_1 + n_2)  \ , \\
p_2 &= n_2 \ .
\label{eq:p_def_2}
\end{align}
This provides a generalized result relating the scaling
of velocity gradients to that of increments in the 
inertial-range. 
Defining 
\begin{align}
p_1 + p_2 = p \ , \quad n_1 + n_2 = n \ ,    
\end{align} 
we can more compactly write $\rho_{n_1, n_2} = p - n$, 
with $p$ being the solution of 
\begin{align}
\zeta_{p - n_2, n_2} + p = 2(n_1 + n_2) \ . 
\label{eq:p_def}
\end{align}

\paragraph{\bf Longitudinal and transverse gradients:}
Consider now the two cases 
1) $n_2=0$ (and $n_1=n$), and 2) $n_1=0$ (and $n_2=n$), 
which isolate the scaling of longitudinal and transverse gradients,
respectively. 
For these cases,  Eq.~\eqref{eq:p_def} yields 
\begin{align}
\label{eq:zeta_l}
\zeta_{p,0} + p &= 2n  \ , \qquad (n_2=0, \ n_1=n) \ , \\ 
\zeta_{p-n,n} + p &= 2n  \ , \qquad (n_1=0, \ n_2=n) \ .
\label{eq:zeta_t}
\end{align}
The first relation in Eq.~\eqref{eq:zeta_l} 
implies that the exponents for
longitudinal gradients $\rho_{n,0}$ are 
prescribed solely by longitudinal scaling exponents
$\zeta_{p,0}$. 
In fact, this result is identical to that obtained from   
conventional univariate multifractal description 
of longitudinal statistics \cite{Nelkin90, Frisch95}.
In striking contrast, 
Eq.~\eqref{eq:zeta_t} reveals that the exponents for transverse
gradients $\rho_{0,n}$ are determined {\em not} by transverse structure
functions $\zeta_{0,p}$, but rather by mixed exponents $\zeta_{p-n, n}$.

It is worth emphasizing that this inherent asymmetry in
scaling of longitudinal and transverse
gradients is rooted in how the dissipative
cutoff $\eta/L$ is defined (in Eq.~\eqref{eq:etah}) solely from
the longitudinal increment.
If the cutoff was instead defined from the transverse increment,
one would obtain the result that scaling of longitudinal gradients
is tied to scaling of mixed-structure functions, while the transverse
statistics are self-prescribed. 
Alternatively, if the dissipative cutoff was defined from a
combination of increments 
\footnote{For instance, this can be done by defining
the cutoff scale using the condition:  
$(\delta u_r)^\alpha  (\delta v_r)^{1-\alpha} r /\nu \simeq 1$,
where $0 < \alpha < 1$ is some parameter that needs to be determined}, 
then both longitudinal and transverse gradients would be tied 
mixed structure function exponents.
However, such considerations can easily be ruled out by appealing
to known results imposed by dissipation anomaly.

For the case $(n_1, n_2) = (2, 0)$, i.e., scaling of 
$\langle (\partial_x u)^2 \rangle$,
Eq.~\eqref{eq:zeta_l} is satisfied
for $p=3$, $\zeta_{3,0}=1$, leading to $\rho_{2,0} = 1$, 
which correspond to the well-known $4/5$-th law,
$\langle (\delta u_r)^3 \rangle = -\frac{4}{5} \langle \epsilon \rangle r$
and dissipation anomaly, $ \nu \langle (\partial_x u)^2 \rangle \sim 1$. 
Likewise, for $(n_1, n_2) = (0, 2)$, i.e., scaling of
$\langle (\partial_x v)^2 \rangle$, 
Eq.~\eqref{eq:zeta_t} is satisfied for $p=3$, 
$\zeta_{1,2}=1$, also leading to $\rho_{0,2} = 1$,
which correspond to the 
$4/3$-th law, 
$\langle \delta u_r (\delta v_r)^2 \rangle = -\frac{4}{3} \langle \epsilon \rangle r$ and dissipation anomaly.
In fact, the results $\zeta_{3,0} = \zeta_{1,2} = 1$
are already well-known to be the only exactly 
derivable results from Navier-Stokes equations,
reflecting a fundamental asymmetry between longitudinal and transverse 
statistics \footnote{Note that from simple symmetry 
arguments, it is
easy to deduce that $\zeta_{0,3}=0$}. 
The relations in Eqs.~\eqref{eq:zeta_l}-\eqref{eq:zeta_t} 
show how this asymmetry propagates to the scaling of velocity gradients and 
generalizes it to higher orders within the joint
multifractal framework.

\paragraph{\bf Validation with DNS results:} 
To assess the validity of
the proposed framework, we next test the
predictions in Eqs.~\eqref{eq:zeta_l}-\eqref{eq:zeta_t} 
with results from DNS. 
The DNS database utilized is   
identical to that in several recent works 
\cite{BPB2020, BBP2020, BS2020, BP2021, BPB2022, BP2023}
and is briefly described in Appendix A.
Before analyzing the scalings, a couple of important considerations should
be noted. The inertial-range exponents for structure 
functions are expected to be independent of $Re$, 
with only the extent of power-laws increasing with $Re$.
In contrast, the scaling of gradient moments explicitly depends
on $Re$, and care must be taken to accurately
extract the exponents at high $Re$. 
In some recent studies \cite{SreeniYakhot:2021, Khurshid_2023}, 
it was argued
that asymptotic power-law scalings for gradient moments
already manifest at low $Re$. 
Compiling data from various sources with that of our own
at higher $Re$, we show in Appendix C that this is not actually the case,
and asymptotic power-laws for gradient moments also only
arise at high $Re$. 

Retaining only the DNS at high enough $\re$, 
the scaling of longitudinal and transverse gradient moments
is shown in Fig.~\ref{fig:mom_all}. 
The central moments as a function of $\re$ are shown,
highlighting small but clear differences between
scaling of longitudinal and transverse moments. 
The expression in Eq.~\eqref{eq:dudxn} can be
modified for central moments defined as 
\begin{align}
M_{n_1, n_2} \equiv  \frac{\langle (\partial_x u)^{n_1} 
(\partial_x v )^{n_2} \rangle} 
{\langle (\partial_x u)^2 \rangle^{n_1/2} 
\langle (\partial_x v )^2 \rangle^{n_2/2}} \sim Re_\lambda^{\xi_{n_1, n_2}} 
\end{align}
\begin{align}
\begin{split}
\text{with} \ , \qquad \xi_{n_1, n_2} & = 2\rho_{n_1, n_2} - n_1 - n_2 \\
               & = 2p - 3n
\end{split}
\end{align}
where we have used $Re \sim Re_\lambda^2$ 
and $\rho_{2,0} = \rho_{0,2} = 1$. 
It is easy to see that $\xi_{2,0} = \xi_{0,2} = 0$,
whereas the results in  
Fig.~\ref{fig:mom_all} provide $\xi_{n,0}$ and $\xi_{0,n}$
for $n=4,6,8$. Additionally, 
the result for skewness of longitudinal gradients,
$\xi_{3,0} \approx 0.14$   is adopted from 
earlier works \cite{Ishihara07, BBP2020, BS2023}.

\begin{figure}
\centering
\includegraphics[width=0.47\textwidth]{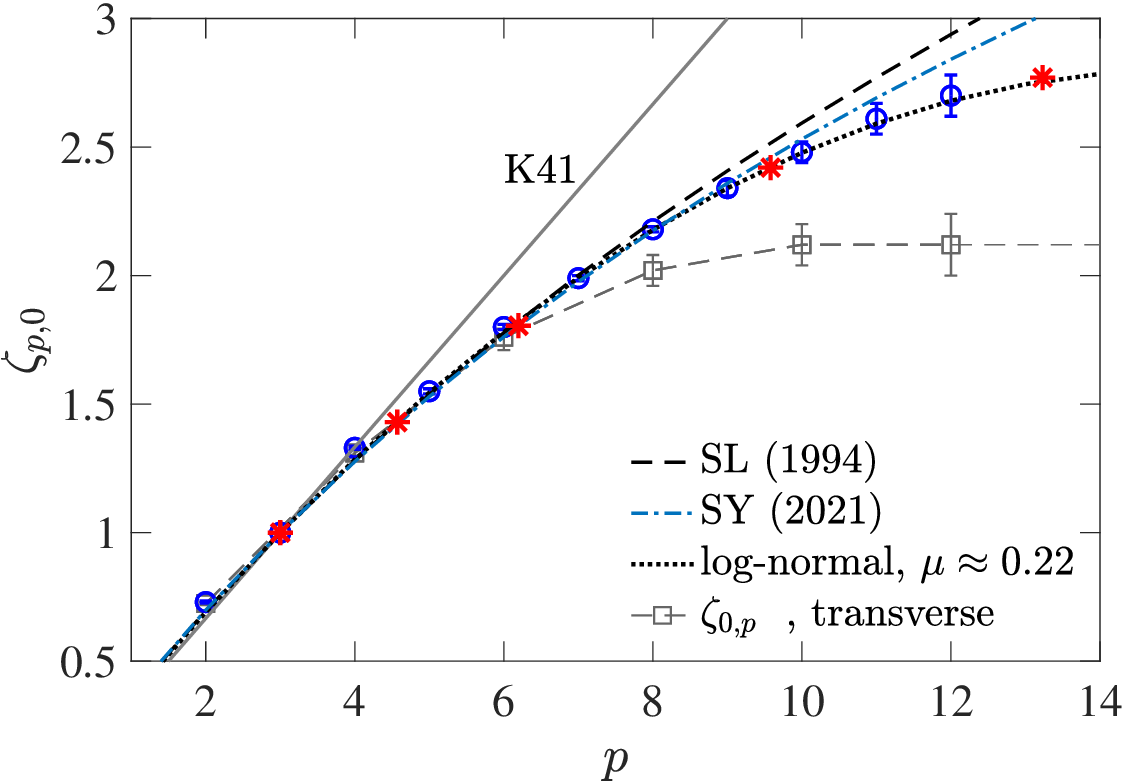} 
\caption{
Longitudinal inertial-range scaling exponents $\zeta_{p,0}$ versus $p$. 
Integer-order exponents (blue circles) are obtained directly from structure functions 
and were reported previously \cite{iyer2020, BS_2023}, while non-integer orders
(red asterisks) are inferred from velocity gradient moments using Eq.~\eqref{eq:long_inv},
for $n=2,3,4,6,8$. For reference, 
theoretical predictions \cite{K62, SL94, SreeniYakhot:2021} and the transverse exponents
$\zeta_{0,p}$ are also shown.
}
\label{fig:zeta_long}
\end{figure}

Focusing first on longitudinal statistics,
to validate the result in Eq.~\eqref{eq:zeta_l},
we simply invert the relation
$\xi_{n,0} = 2p- 3n$ and
Eq.~\eqref{eq:zeta_l} to solve for 
for $p$ and $\zeta_{p,0}$, leading to
\begin{align}
p = (3n + \xi_{n,0})/2 \ , \qquad
 \zeta_{p,0} = 2n-p \ .
 \label{eq:long_inv}
\end{align}
It is easy to see that 
these relations
imply  $3n/2 < p < 2n$, suggesting
that $n$-th order gradient arise
from structure functions of order
greater than $3n/2$ but smaller than $2n$. 
Now for various $n$ and $\xi_{n,0}$ 
from Fig.~\ref{fig:mom_all}, we can obtain pairs
of $p$ and $\zeta_{p,0}$, and compare them with known
results for longitudinal inertial-range exponents 
directly obtained from structure functions.
This comparison is shown in
Fig.~\ref{fig:zeta_long},
with exponents for integer $p$ values
obtained from recent works \cite{iyer2020, BS_2023}. 
and non-integer $p$ values corresponding
to scaling exponents of longitudinal gradients
for $n=2,3,4,6,8$. 
Evidently, remarkable agreement is obtained
between them, with all exponents
near-perfectly described by the
log-normal prediction 
(with intermittency exponent $\mu \approx 0.22$
\cite{BS2022}.)

Is it well-known that the log-normal prediction is untenable
for large $p$ values \cite{Frisch95},
nevertheless, it serves as a remarkable approximation of the exponents
up to orders shown in Fig.~\ref{fig:zeta_long}. 
For comparison, we also show the theoretical predictions
of \cite{SL94, SreeniYakhot:2021}, which show conspicuous
deviations for large $p$. For this reason, any predictions
for gradient exponents even at modest orders
obtained from these results would deviate from DNS results,
as was the case in \cite{elsinga2023}. 
While a fully accurate theoretical prediction for $\zeta_{p,0}$ for highest
orders remains elusive, it is evident from Fig.~\ref{fig:zeta_long},
that scaling of structure functions and gradient moments
as described by the joint multifractal framework is fully self-consistent.

\begin{figure}
\centering
\includegraphics[width=0.47\textwidth]{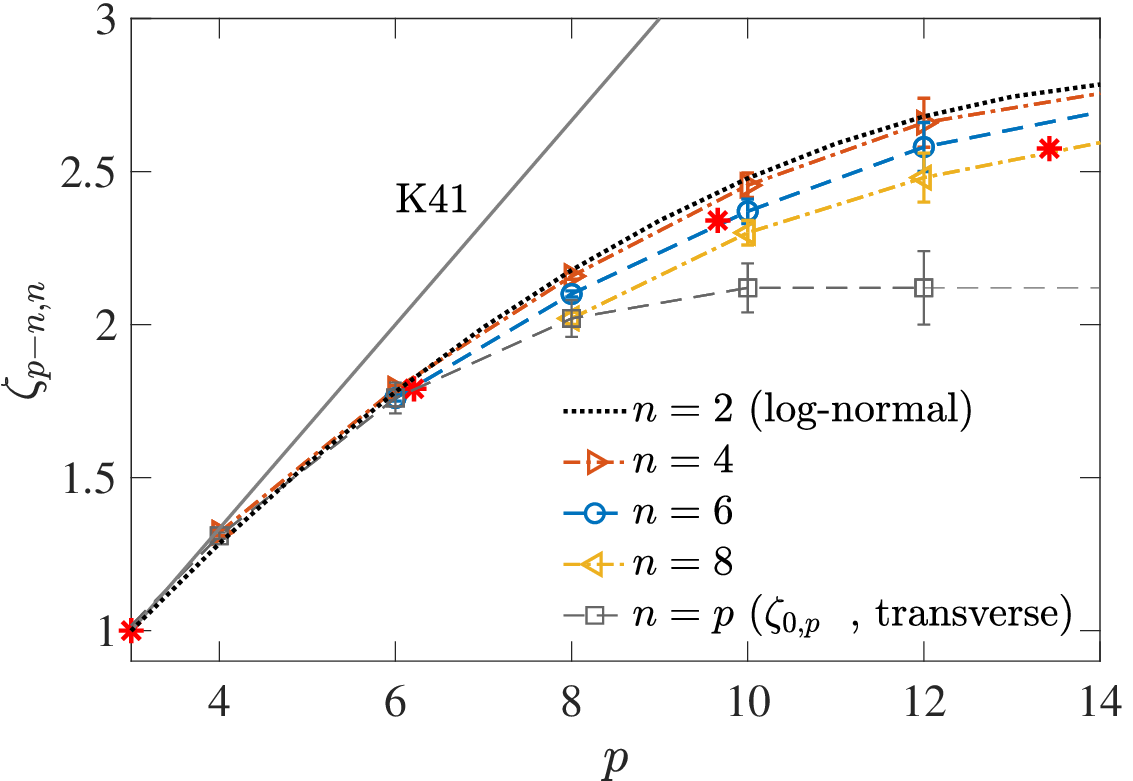} 
\caption{
Mixed inertial-range scaling exponents $\zeta_{p-n,n}$ versus $p$ 
for various integer values of $n$ (see legend). Exponents at integer 
$p$ are obtained directly from structure functions, while non-integer 
orders (red asterisks) are inferred from the scaling of transverse velocity gradient 
moments using Eq.~\eqref{eq:tran_inv} for $n=2,4,6,8$. The dashed curve for $n=2$ 
corresponds to the log-normal prediction from Fig.~\ref{fig:zeta_long}
and represents $\zeta_{p-2,2}$ 
(which is identical to $\zeta_{p,0}$ \cite{SreeniYakhot:2021}). 
}
\label{fig:zeta_tran}
\end{figure}

Next, focusing on transverse gradients, 
we can invert 
$\xi_{0,n} = 2p- 3n$ and
Eq.~\eqref{eq:zeta_t} to write
\begin{align}
p = (3n + \xi_{0,n})/2 \ , \qquad
\zeta_{p-n,n} = 2n-p \ .    
 \label{eq:tran_inv}
\end{align}
Thus, in this case, each gradient moment
of order $n$, falls on a different branch
of mixed structure function prescribed by
$\zeta_{p-n,n}$. To that end, we
evaluate these exponents for $n=2,4,6,8$
from $\xi_{0,n}$, and simultaneously
extract $\zeta_{p-n, n}$ directly from 
mixed structure functions in the inertial-range
(the latter by extending the procedures in
\cite{iyer2020, BS_2023}). 
These results are shown in 
Fig.~\ref{fig:zeta_tran}. 
Incidentally, we find that $\zeta_{p-2, 2} \approx \zeta_{p,0}$,
for $p\geq2$, which has been theoretically 
suggested before \cite{SreeniYakhot:2021}.
Remarkably, we observe that the
results for $p \ , \zeta_{p-n,n}$ obtained from
gradient exponents fall near perfectly
on different branches of $\zeta_{p-n,n}$ 
obtained from structure functions for integer
values of $p$.

\paragraph{\bf Conclusions:}

To summarize, we have developed here a unified multifractal
framework that jointly describes longitudinal and transverse
intermittency in fully developed turbulence, 
naturally linking intermittency of velocity increments in the inertial-range
and that of gradients at the dissipative scales. 
Our analysis reaffirms that longitudinal gradient
statistics are prescribed solely by longitudinal
structure functions, however, transverse gradient statistics
are governed by mixed structure functions. 
The resulting predictions show near-perfect agreement with DNS
data, establishing both the consistency and accuracy of the framework.

Our results clearly emphasize that a complete and faithful description
of intermittency requires simultaneous consideration
of longitudinal and transverse fluctuations,
and hence intrinsically coupled dynamics
of stretching and rotational motions encoded in Navier-Stokes
equations. While recent studies have also hinted at this
\cite{BPBY2019, Carbone:20b, johnson2020, BP2022, BP2023},
the present work now provides an explicit statistical framework
for it.
Although the framework establishes precise connections between inertial
and dissipation-range statistics, it does not yet furnish a
closed description for the full set of 
mixed exponents $\zeta_{p_1, p_2}$,
which remains an outstanding challenge. 
Addressing this gap will surely require development
of  intermittency models that embrace the tensorial nature
of underlying turbulence dynamics. 

The unified description developed
here also carries potential implications for turbulence modeling.
By explicitly accounting for the coupled nature of longitudinal
and transverse intermittency, it offers a natural
pathway towards improved representation of small-scales
in reduced-order models and large-eddy simulations. 
Additionally, the framework also opens several promising directions
for future investigation,
including extensions to Lagrangian intermittency--
especially acceleration statistics, which 
have proven particularly challenging to characterize 
under the multifractal framework \cite{BS_PRL_2022, BS2023}. 
Progress along these directions will be reported as future work.

\begin{acknowledgements}
\paragraph*{\bf Acknowledgments:}
We gratefully acknowledge the Gauss Centre for Supercomputing 
e.V. (www.gauss-center.eu) for providing time on the supercomputers 
JUQUEEN and JUWELS at J\"ulich Supercomputing Centre (JSC),
where the simulations reported in this
paper were performed. We are also grateful to Sualeh Khurshid
and Toshiyuki Gotoh for sharing the data points used
in Fig.~\ref{fig:mom4}.
\end{acknowledgements}

\clearpage

\subsection*{Appendix A -- DNS data}

The DNS data utilized here are obtained by solving the 
incompressible Navier-Stokes equations:
\begin{equation}
\partial \uu / \partial t + ( \uu \cdot \nabla) \uu = - \nabla P + \nu \nabla^2 \uu + \mathbf{f} \label{eq:NS}  \, ,
\end{equation}
where $\uu$ is the divergence-free velocity $(\nabla \cdot \uu = 0)$, $P$ is the 
kinematic pressure, 
and $\mathbf{f}$ is the forcing term. 
The simulations correspond to the canonical setup of forced
isotropic turbulence in a cubic domain
with periodic boundary conditions.
This allows us to use
highly accurate Fourier pseudo-spectral methods, 
with aliasing error controlled by a combination of
truncation and phase-shifting \cite{PattOrs71,  Rogallo}.
The largest scales are forced stochastically to maintain a 
statistically stationary state \cite{EP88}.  
The domain size is $(2 \pi)^3$, discretized into $N^3$ grid points, 
with uniform grid-spacing $\Delta x = 2\pi/N$ in each direction. 
The Taylor-scale Reynolds number $\re$ goes from $140$ to $1300$ -- 
on grid sizes $1024^3$ to $12288^3$,  respectively, thus maintaining
excellent small scale resolution \cite{BPBY2019, BBP2020}.
As mentioned earlier, the precise data have been used in numerous recent
studies \cite{BBP2020, BPB2020, BS2020, BP2021, BPB2022, BP2023},
which have adequately established convergence
with respect to resolution and statistical sampling. 
In fact, in some recent studies \cite{BLW:2024, BP:2025}, 
detailed comparisons have also been made
with laboratory experiments and DNS of some anisotropic flows, 
showing excellent
quantitative agreement.

\subsection*{Appendix B -- Derivation of Eqs.~\eqref{eq:rho_def}-\eqref{eq:p_def_2}}

For the result in Eq.~\eqref{eq:rhon}, defining the critical 
H\"older exponents as $h_1^*$ and $h_2^*$, the right-hand side
yields the following after taking the derivatives
\begin{align}
\begin{split}
(1&+h_1^*)(-n_1 + \partial D/\partial h_1 )  \\ 
&- [n_1 (1 - h_1^*) + n_2 (1-h_2^*) - 3 + D] = 0  \  ,
\end{split} 
\label{eq:rdev1}
\end{align}
\vspace{-0.5cm}
\begin{align}
-n_2 + \partial D/\partial h_2 &= 0 \ .
\end{align}
Substituting $\partial D/\partial h_1 = p_1$, 
$\partial D/\partial h_1 = p_2$ from Eq.~\eqref{eq:hcrit}
and $-3 + D = p_1 h_1^* + p_2 h_2^* - \zeta_{p_1, p_2}$
from Eq.~\eqref{eq:zetap} then yields
\begin{align}
\zeta_{p_1, p_2} + p_1 &= 2n_1 + n_2 + (p_2 - n_2) h_2^* \ , \\
 p_2 &= n_2 \ , 
\end{align}
which can be readily simplified to give the results
in Eqs.~\eqref{eq:p_def_1}-\eqref{eq:p_def_2}. 
The result in Eq.~\eqref{eq:rho_def} can be obtained by 
combining the result in Eq.~\eqref{eq:rdev1} 
with that in Eq.~\eqref{eq:rhon}.

\begin{figure}
\centering
\includegraphics[width=0.45\textwidth]{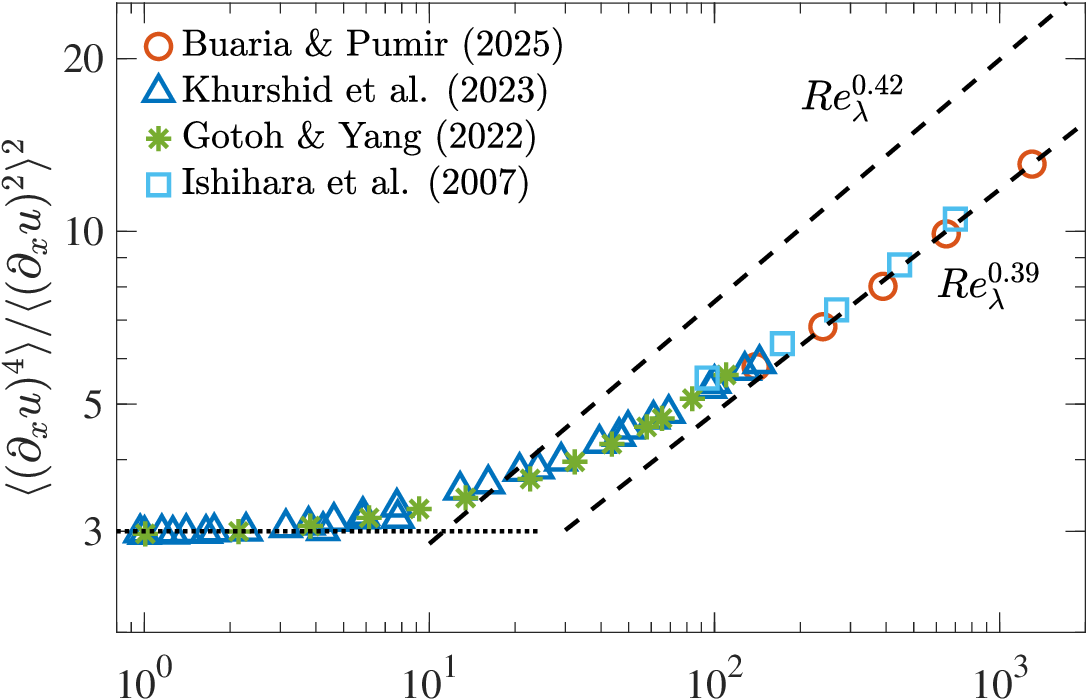} \\ 
\vspace{0.25cm}
\includegraphics[width=0.45\textwidth]{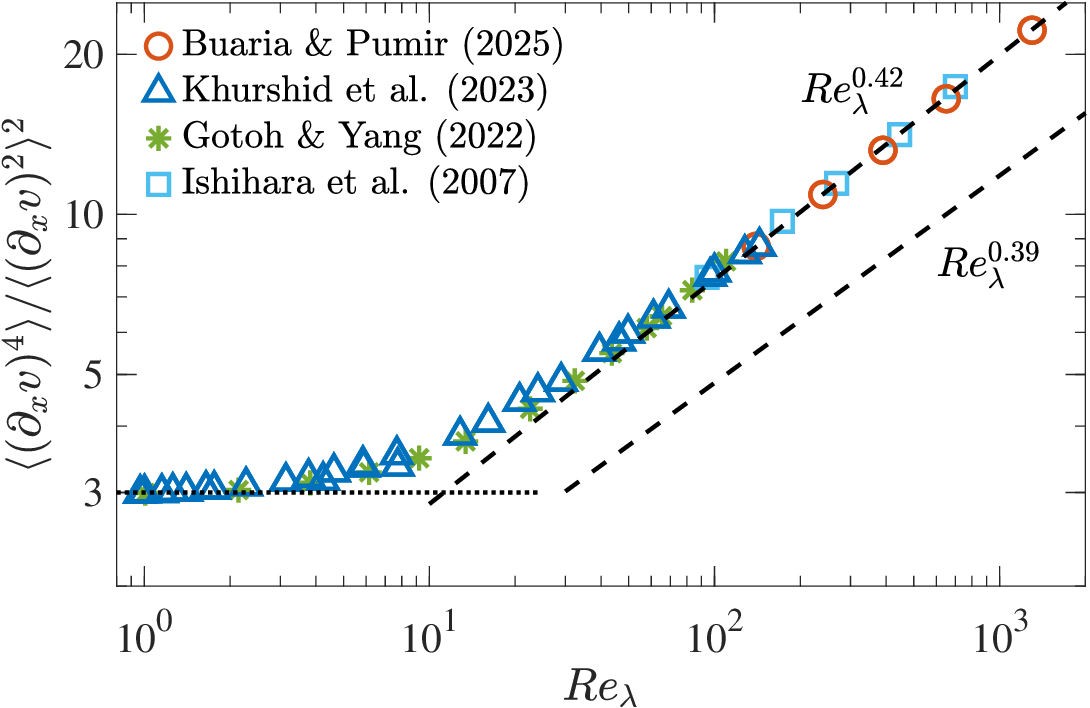} 
\caption{
Flatness of longitudinal (top) and transverse (bottom) velocity gradients,
compiled using data from various sources 
\cite{Ishihara07, Gotoh:2022, Khurshid_2023, BP:2025}.
The best-fit power-law scaling are also marked in dashed lines.
The dotted horizontal line at $3$ marks the flatness
of a standard Gaussian distribution. 
}
\label{fig:mom4}
\end{figure}

\subsection*{Appendix C -- Asymptotic scaling of gradient moments}

Recently, there has been an some debate as to when precisely
does the asymptotic scaling of velocity gradient moments
emerges as a function of $\re$. 
When considering inertial-range scaling of structure functions,
the scaling exponents emerge as plateaus on local slope
plots, with only the extent of the plateau increasing
with $\re$, but the value itself being independent of $\re$.
See e.g. \cite{iyer2020, BS_2023} for results
at the $\re$ considered in this work.
The use of extended self-similarity \cite{Benzi+93}
can often provide a more robust plateau, especially at lower $\re$. 
In contrast, the situation for gradient moments
is tricky, since the scaling exponents have to be 
directly extracted as a function of $\re$. 
Classical theoretical arguments dictate that scaling exponents
for gradients should also require high $\re$.
However, some recent studies 
\cite{Schumacher+14, SreeniYakhot:2021, Khurshid_2023}
have suggested that asymptotic scaling of gradient moments
emerges at very low $\re$ ($\gtrsim 10$), substantially
lower than dictated by theory. 
However, these studies were all restricted to $\re \lesssim 200$,
and therefore did not verify if the same scaling behaviors extend to much
higher $\re$, as considered here.

To resolve this, we have compiled
data from DNS of isotropic turbulence
from numerous sources \cite{Ishihara07, Gotoh:2022, Khurshid_2023}
along with our own \cite{BP:2025}. 
Figure~\ref{fig:mom4} shows the flatness of longitudinal
and transverse velocity gradients, 
spanning three orders of magnitude
in $\re$ (and six in $Re$), providing a comprehensive
view of the scaling behavior.  Excellent correspondence is obtained
between all the data. 
Several important observations can be made
from these plots. At very low $\re$, the flatness starts at
$3$, which corresponds to a Gaussian distribution. 
With increasing $\re$, the flatness steadily increases
and approaches a clear asymptotic power-law only at the 
highest $\re$ (which are very slightly different for longitudinal
and transverse gradients). 
The intermediate range, in particular $10 \lesssim\re \lesssim 200$
clearly corresponds to some kind of smooth transition regime,
where turbulence has not fully developed yet. 
Thus, any power-laws extracted in this regime, do not truly
correspond to asymptotic scalings of gradient moments --
contrary to what was proposed in earlier works
\cite{Schumacher+14, SreeniYakhot:2021, Khurshid_2023}. 
Although not shown, 
same conclusions can also be drawn from the behavior of
higher-order moments. 

Interestingly, the work of \cite{SreeniYakhot:2021}, SY2021 henceforth,
also provides a way to extract scaling of gradient moments
from inertial range exponents. However, unlike the multifractal
results derived in this work, their theory posits
that gradient moments of order $n$ directly result
from structure functions of order $p=2n$.
The corresponding result in multifractal model gives
$3n/2 < p < 2n$, resulting in stronger scaling laws 
than predicted by SY2021. 
In SY2021, the power-laws for gradient moments were
extracted at low $\re$, resulting in fortuitous agreement
between the predictions from their theory. However, given the 
data at much higher $\re$ and the present theoretical developments, 
it stands to reason that the multifractal extension of inertial-range
intermittency to gradients provides the correct result.

On a related note, the recent work of \cite{elsinga2023}
has suggested that the true asymptotic scalings of gradients might still not have been
realized, with exponents
still increasing with $\re$, and  longitudinal
and transverse exponents potentially becoming identical at extremely high $\re$.
This certainly cannot be ruled out given the highest $\re$ in DNS
are still modest compared to what can be realized in
geophysical  flows. 
However, the trends in Fig.~\ref{fig:mom4} definitely suggest 
any further change to the exponents, if happening, 
would be extremely slow.
In fact, a simple extrapolation of current data trends
would show that such a change 
might only become perceptible
at $\re \gtrsim 10,000$ (or even higher), well beyond what can be realized in 
DNS or laboratory experiments for the foreseeable future.
Moreover, any change in scaling of gradients would also be 
accompanied by change in inertial-range exponents.
These aspects certainly need to kept in mind for future,
but given the available data, 
the proposed joint multifractal framework appears to be
self-consistent and accurate in explaining the scaling of 
longitudinal and transverse gradients.

\end{document}